# Silicon-on-insulator optomechanical microresonator with tight photon and phonon confinement

J. Zhang[1,*], X. Le Roux[1], M. Montesinos-Ballester[1], O. Ortiz[1], D. Marris-Morini[1], L. Vivien[1], N. D. Lanzillotti-Kimura[1], and C. Alonso-Ramos[1]

[1] Université Paris-Saclay, CNRS, Centre de Nanosciences et de Nanotechnologies, 91120 Palaiseau, France

jianhao.zhang@c2n.upsaclay.fr

The implementation of optomechanical devices in silicon-on-insulator (SOI), the canonical silicon photonics technology is seriously hampered by the strong phonon leakage into the silica under-cladding. This limitation has been partially circumvented by total or partial removal of the silica under-cladding to form Si membranes or pedestal waveguides. However, this approach complicates integration with standard silicon optoelectronics circuitry, limiting the versatility and application of the strategy. Here, we propose and demonstrate a new strategy to confine photons and phonons in SOI without removing the silica under-cladding. Inspired by end-fire antenna arrays, we implement a periodic nanostructuration of silicon that simultaneously enables cancelling phonon leakage by destructive interference and guiding of photons by metamaterial index confinement. Based on this concept, we implement SOI optomechanical micro-resonators yielding remarkable optomechanical coupling ($g_o$=49 kHz) between 0.66 GHz mechanical modes and near-infrared optical modes. The mechanical mode exhibits a measured quality factor of $Q_m$ = 730, the largest reported for SOI optomechanical resonators, without silica removal. This value compares favorably with state-of-the-art Si membrane waveguides recently used to demonstrate remarkable Brillouin interactions in silicon ($Q_m \sim 700$). These results open a new path for developing optomechanics in SOI without the need for silica removal, allowing seamless co-integration with current Si optoelectronics circuits, with a great potential for applications in communications, sensing, metrology, and quantum technologies.

**Introduction**

The coupling of mechanical and optical waves via radiation pressure has been a subject of early research in the context of bistable optical interferometers [1] and studied intensively in the field of cavity optomechanics [2, 3]. Being one particular optomechanical interaction, Brillouin scattering [4] entails the coupling [5, 6] of optical photons and acoustic phonons with frequencies ranging between a few tens of MHz and hundreds of GHz. Brillouin scattering is of tremendous interest to integrated photonics [7,8] given its singular optomechanical applications spanning laser physics, microwave-frequency phonon excitation, and precision spectroscopy, among others. To date, on-chip Brillouin optomechanics [9] has been implemented using chalcogenide-based materials [10], gallium arsenide [11-14], and silicon [15-18]. Implementation of Brillouin interactions in silicon-on-insulator, the preferred silicon photonics technology, remains an open challenge. Photons are confined in dense media with a high refractive index, while phonons are confined in soft and light materials. Silicon has a higher refractive index and is stiffer than silica. Thus, SOI waveguides yield tight photon confinement and strong phonon leakage towards the silica under-cladding [6, 8, 9]. This limitation has been circumvented by partial [16] or total [17-22] removal of the silica under-cladding. Most remarkable results have been obtained using Si rib membrane waveguides [17], showing lasing [18] and non-reciprocal modulation [22], among others. Nevertheless, the need for silica under-cladding removal hampers the co-integration of these devices with Si optoelectronics, limiting the applicability of this approach. Initial attempts to simultaneously confine photons and phonons in SOI waveguides have been recently reported relying on geometrical softening. That is, the lowering silicon stiffness by reducing the width of the waveguide. However, narrowing down the waveguide hinders optical confinement. Slot waveguides, comprising two parallel narrow Si strips, have been proposed to reduce Si stiffness while providing optical guiding [23, 24]. However, phase-matching conditions for optomechanical backaction impose operation in the leaky wave regime, seriously compromising the mechanical quality factor ($Q_m$~100).

Here, we propose a radically different approach to confine photons and phonons in SOI without removing silica under-cladding. The idea is to periodically pattern silicon waveguides, harnessing dipole-dipole interactions to cancel phonon leakage by destructive interference while exploiting optical metamaterial engineering to confine and guide photons. Based on this concept, we demonstrate SOI optomechanical micro-resonators with mechanical quality factors as high as $Q_m$ = 730.

## Results

### Phonon and photon confinement in nanostructured SOI waveguides

Figure 1(a) shows a schematic view of the proposed nanostructured SOI optomechanical waveguide. First, we discuss the suppression of phonon leakage towards the silica under-cladding. Interactions among pairs and arrays of semiconducting micropillars have been studied, considering strain coupling through common substrate [25], and hybridization of the resonating pillars with the continuum of elastic waves of the substrate [26]. Here, we exploit destructive interference to suppress phonon leakage towards the silica under-cladding. As shown in Fig. 1(b), the mechanical mode of an isolated silicon pillar has strong leakage into the silica under-cladding. This leakage can be described as the radiation of an antenna, exhibiting a quasi-omnidirectional distribution (see Figs 1(c) and 1(d)). In a similar fashion, the proposed optomechanical waveguide, comprising a periodic arrangement of silicon pillars, can be described as an antenna array. The radiation characteristics of an antenna array comprising a periodic arrangement of omnidirectional antennas are governed by the pitch, Λ in Fig. 1(a), and the relative phase between the antennas [27]. When the pitch is half of the wavelength, and the excitation of adjacent antennas is shifted by $\pi$, the radiation perpendicular to the antenna array is cancelled. Here, we exploit this array configuration, usually referred to as end-fire, to suppress phonon leakage (radiation) towards the silica under-cladding.

We set a period of Λ = 300 nm, equal to half the mechanical wavelength (~ 600 nm), and a gap length of $L_g$ = 100 nm. The pillar thickness is t = 500 nm and the width is W = 800 nm. We calculate the mechanical modes of the periodic waveguide, considering a unit cell comprising two pillars. We find two mechanical modes with in-phase (Fig. 1(e)) and $\pi$-shifted (Fig. 1(h)) motions, respectively. As shown in Figs. 1(f) and 1(g), the mode with in-phase motion presents strong radiation towards the silica cladding. Conversely, the mechanical mode with $\pi$-shifted motion exhibits very weak radiation towards the silica (Figs. 1(i)-1(j)), with negligible values in the direction perpendicular to the waveguide array. This radiation pattern matches very well with that of an antenna array in an end-fire configuration [27].

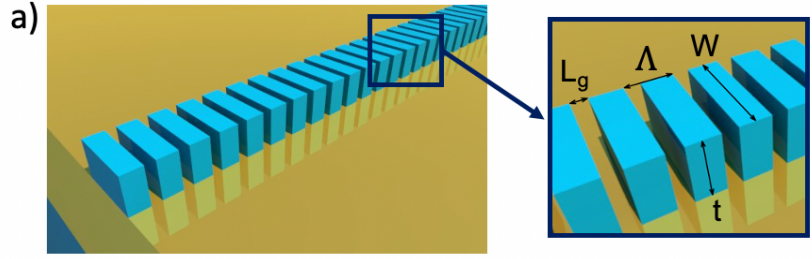

- Single silicon pillar

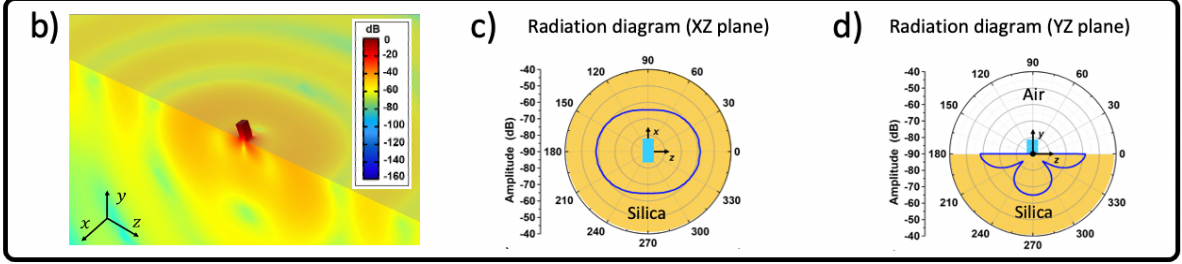

- Silicon pillar array: in-phase motion

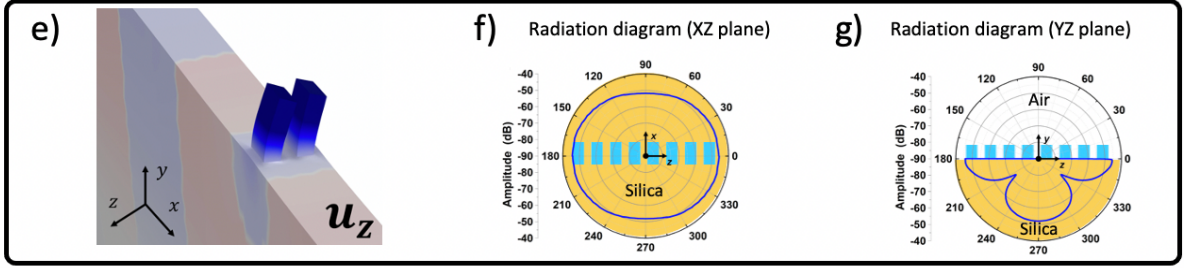

- Silicon pillar array: $\pi$-shifted motion

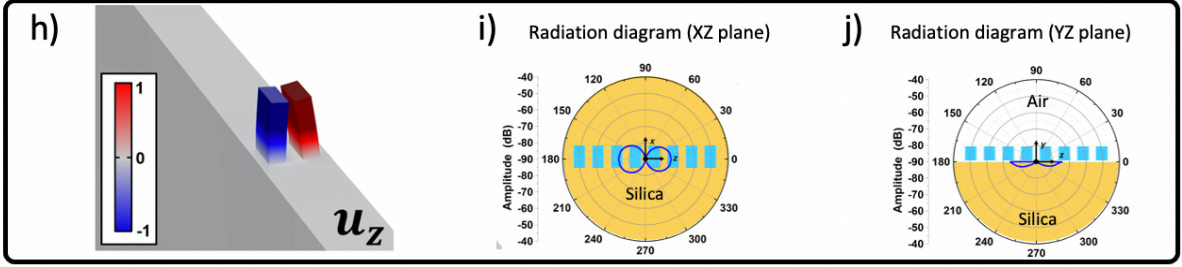

**Figure 1:** a) Schematic of the proposed periodic optomechanical waveguide in SOI. Mechanical mode in isolated silicon pillar behaving like a quasi-omnidirectional radiator in the silica under-cladding: b) normalized displacement profile, radiation diagram in c) horizontal XZ plane, and d) vertical YZ plane. Mechanical mode in the proposed optomechanical waveguide with in-phase vibration of adjacent silicon pillars (broadside like): e) normalized displacement profile, radiation diagram in f) horizontal XZ plane and g) vertical YZ plane. Mechanical mode in the proposed optomechanical waveguide with contra-phase vibration of adjacent silicon pillars (end-fire like): h) normalized displacement profile, radiation diagram in i) horizontal XZ plane and j) vertical YZ plane.

We calculate the quality factor of the mechanical modes ($Q_m$) of the periodic waveguide, considering the thermo-elastic loss ($Q_{TE}$) and the losses from the acoustic leakage ($Q_L$) towards the silica under-cladding. The quality factor of the mechanical mode $Q_m$ can therefore be considered via

$$\frac{1}{Q_m} = \frac{1}{Q_{TE}} + \frac{1}{Q_L} \qquad (1)$$

We use a silicon-on-insulator stack with a 500-nm-thick top silicon layer, 5-µm-thick silica under-cladding, and infinite silicon substrate. In Fig. 2(a), we show mechanical quality factor and frequency, calculated as a function of the period, $\Lambda$, for the two mechanical modes: in-phase motion and $\pi$-shifted motion (end-fire-like). The mechanical quality factor of the $\pi$-shifted mode reaches a maximum for a period of 300 nm (equal to half of the wavelength) when end-fire conditions are satisfied [27]. At the optimum period ($\Lambda$ = 300 nm), the quality factor of the $\pi$-shifted mode, ~2 ×10$^4$, is four orders of magnitude larger than that of the mode with in-phase motion. Figure 2(b) shows the calculated displacement profile of the $\pi$-shifted mode for $\Lambda$ = 300 nm. The energy is greatly confined in the silicon pillars and the very surface of the silica under-cladding, illustrating the effective phonon confinement achieved with our strategy.

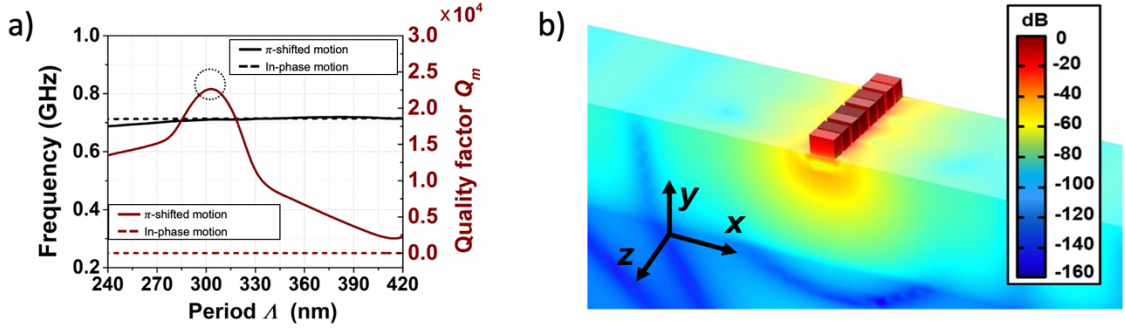

**Figure 2:** a) Mechanical quality factor, $Q_m$, and frequency calculated as a function of the period for the two modes (in-phase, counter-phase) in the proposed SOI optomechanical waveguide. b) Normalized displacement profile of counter-phase (end-fire-like) mechanical mode of the proposed optomechanical waveguide.

From the optical point of view, the proposed nanostructured Si waveguide is a conventional 1D photonic crystal. Figure 3(a) shows the optical band structure for the transverse electric (TE) polarized mode. Below the bandgap, in the metamaterial regime [28, 29], the periodic waveguide supports a well-confined optical mode (see Fig. 3(b)). Interestingly, the optical mode has a remarkably strong electrical field in the longitudinal direction ($E_z$). The motion of the mechanical modes occurs mainly along the longitudinal direction, z. Thus, the strong $E_z$ optical field favors the optomechanical backaction [2, 3].

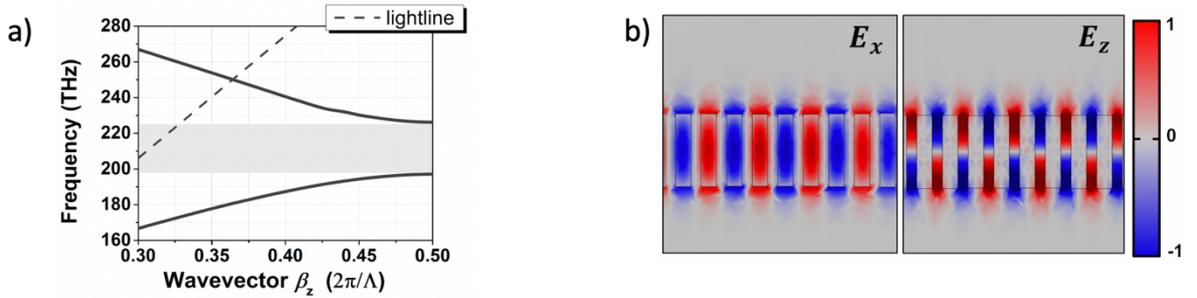

**Figure 3:** a) optical band diagram of the proposed nanostructured waveguide. b) Transversal ($E_x$) and longitudinal ($E_z$) electrical field distributions of the TE mode at 1550 nm wavelength.

## SOI nanostructured optomechanical microresonator

To demonstrate the proposed concept, we developed an optomechanical micro-resonator in the SOI technology. Figure 4(a) shows a schematic view of our 1D photonic crystal microresonator. The cavity is formed by parabolic narrowing of the waveguide width towards the center, from $W_m$ = 1000 nm to $W_0$ = 800 nm. We keep a constant period ($\Lambda$) and gap length ($L_g$). This approach yields optical modes with high-quality factors [30] while maintaining the phonon leakage suppression. The optical profiles of the first 3 optical resonant modes are shown in Fig. 4(b). The calculated quality factors of these 3 modes are $1.5 \times 10^5$, $6 \times 10^4$, and

$1 \times 10^4$, respectively. The proposed optomechanical resonator supports more than 20 resonant mechanical modes with high-quality factors ($Q_m > 1 \times 10^4$). As an illustration, we show the displacement profile of three mechanical modes in Fig. 4(c).

To best excite the $\pi$-shifted mechanical modes, we choose the first order, asymmetric, optical mode. To evaluate the optomechanical interaction [2, 3, 31-34], we use the single-photon optomechanical coupling rate $g_0$ which describes the coupling strength between a photon and a phonon in the optomechanical system [2, 3, 5]. As shown in Fig. 4(d), the first-order optical mode yields a calculated coupling rate exceeding 400 kHz for several mechanical modes near 0.67 GHz.

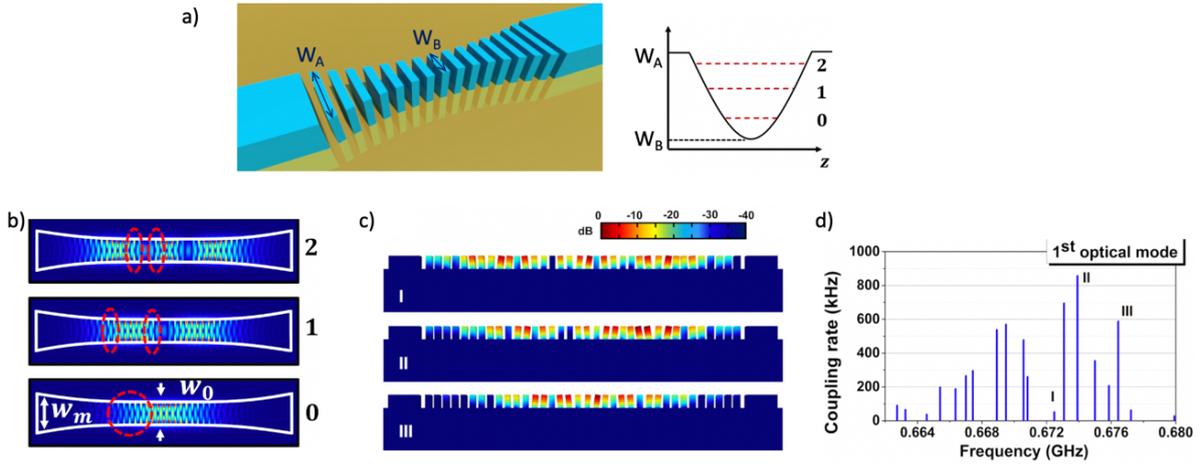

**Figure 4:** a) Schematic view of proposed optomechanical micro-resonator with parabolic width profile. b) Field distribution of the first three optical micro-resonator modes. c) The cut-plane displacement profile of three mechanical micro-resonator modes. d) Optomechanical coupling rate between the first-order optical mode and different mechanical modes in the micro-resonator

We fabricated the optomechanical resonator using standard silicon photonics process, including electron-beam lithography and reactive ion etching. Figure 5(a) shows scanning electron microscope (SEM) images of the optomechanical resonator.

The optomechanical backaction is probed using the setup described in Fig. 5 (b). The optical reflection and transmission are monitored with an optical/electronic spectrum analyzer (OSA/ESA). Experimental measurements were performed at room temperature and at atmospheric pressure. The first-order optical mode has a wavelength of 1560 nm and a measured quality factor of $4.1 \times 10^4$. This optical resonance is excited using a tunable laser and erbium-doped fiber amplifier, yielding 1 mW coupled into the access silicon waveguide. The pump wavelength is then scanned to approach the resonance from a blue-detuned position. This detuning stiffens the optomechanical resonator [2, 3]. Due to thermo-optic and nonlinear effects in silicon, the cavity resonance drifts dynamically with the increasing intra-cavity power. Figure 5(c) shows the optical transmittance, normalized to the on-resonance transmission, and the estimated intracavity power for different wavelengths approaching the resonance. The gray stripe indicates the wavelength of the cold cavity. The dynamics of optomechanical backaction are presented in Fig. 5(d), showing a clear signature of several mechanical modes for intra-cavity optical power exceeding 33 mW. We can observe 9 mechanical resonances near 0.66 GHz. The direct mapping between calculated (Fig. 4(d)) and measured acoustic modes is challenging (Fig. 5(b)). Nevertheless, simulations and experiments show a reasonably good agreement with similar closely-packed fingerprints of several mechanical modes near 0.66 GHz. From the frequency dynamics and the intra-cavity power, we estimate a coupling rate [2, 3, 5, 9] of 49 kHz. Four of the mechanical modes exhibit a mechanical quality factor exceeding 700. This is the highest value demonstrated of an SOI optomechanical resonator without silica under-cladding removal. Furthermore, this mechanical quality factor compares with that of Si rib membranes used for state-of-the-art demonstration of Brillouin functionalities. These results demonstrate the effectiveness of the phonon leakage cancellation approach and its potential for strong backaction between highly confined optical and mechanical modes.

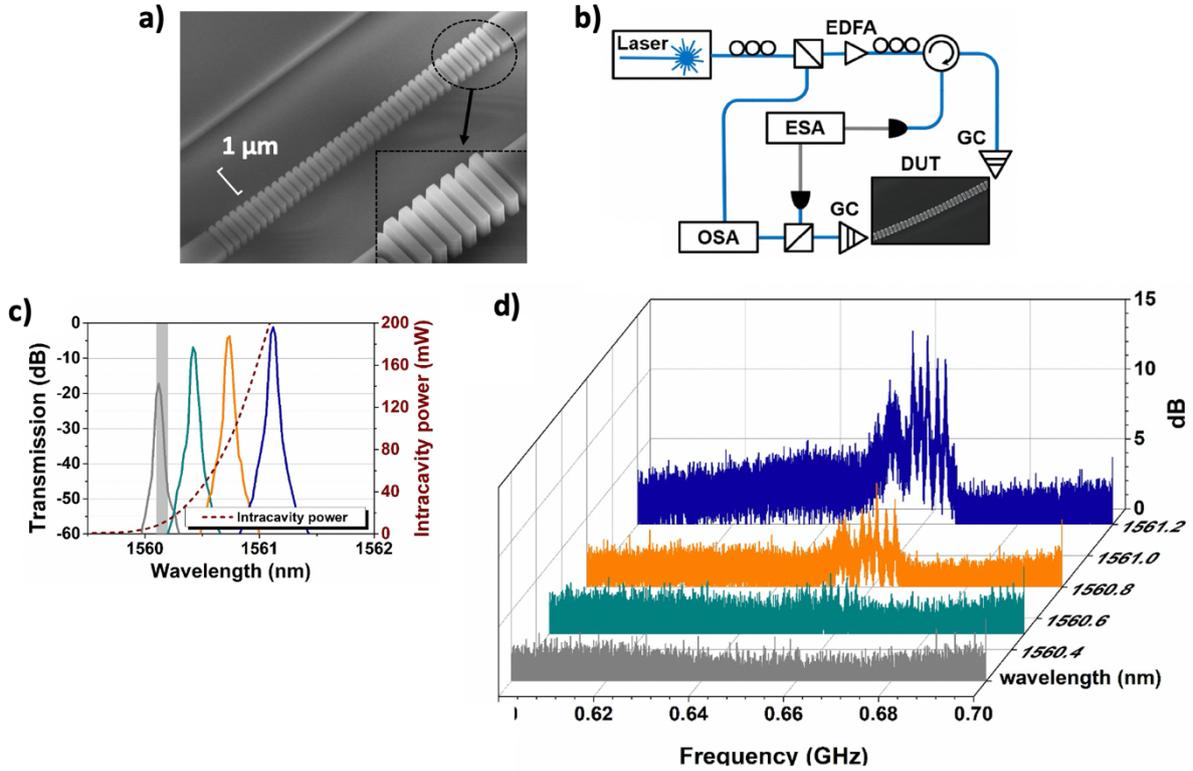

**Figure 5:** a) Scanning electron microscope images of the fabricated optomechanical micro-resonator. b) The setup used for optomechanical characterization. ESA: electronic spectrum analyzer. OSA: optical spectrum analyzer. GC: grating coupler. DUT: device under test. c) Optical dynamics of the resonator when the wavelength scanned over the resonance. The grey stripe indicates the position of the cold resonance. d) Evolution of the radio-frequency spectrum when the pump laser is scanned over the optical resonance. The optical power in the bus waveguide is fixed at 1 mW.

## Discussion

In summary, we have proposed and demonstrated a new strategy to confine photons and phonons in SOI without silica under-cladding removal. This strategy provides a dramatic reduction of the phonon leakage towards the silica under-cladding, overcoming one of the major limitations of optomechanics in SOI. State-of-the-art silicon optomechanical devices circumvent phonon leakage by totally or partially removing the silica under-cladding to form a pedestal or membrane waveguides. This approach allowed the demonstration of a myriad of high-performance optomechanical devices in silicon. However, the need for silica removal hinders co-integration with standard optoelectronic circuitry, offsetting one of the key advantages of silicon photonics.

The innovative strategy is to exploit the nanostructuration of SOI waveguides to suppress phonon leakage by destructive interference. Inspired by radiation control in end-fire antenna arrays, we engineer the periodic nanostructuration, properly combining a series of leaky silicon nanopillars to yield a well-guided mechanical mode. Concurrently, the periodic waveguide acts as a 1D photonic crystal for the optical mode, allowing metamaterial guiding and control of the field distribution, instrumental for strong optomechanical backaction. Based on this approach, we implement optomechanical resonators that harness mechanical leakage suppression and optical metamaterial engineering to yield tightly confined optical and mechanical modes with high-quality factors. We experimentally characterize our SOI optomechanical microresonator at room temperature and ambient conditions, showing optical quality factor of $4 \times 10^4$, near 1560 nm wavelength, mechanical quality factors up to 730, near 0.66 GHz, and optomechanical coupling rate of 49 kHz. This is the largest mechanical quality factor for an SOI optomechanical resonator without silica under-cladding removal. Although still far from mechanical quality factors in membrane phononic and phoxonic crystals [20, 21], usually in the range of 2000-3000, the value reported here compares with state-of-the-art pedestal and Si rib membranes recently used to demonstrate outstanding Brillouin features like lasing or non-reciprocal modulation.

By combining concepts from antenna theory, metamaterial engineering, and nano-optomechanics to overcome the phonon leakage challenge, the strategy proposed here opens a new path to exploit optomechanical interactions in SOI, the canonical silicon photonics technology. Our approach has a great potential for seamless co-integration of high-performance optomechanical and optoelectronic devices in a single silicon chip. We foresee that our results will expedite the development of a new generation of optomechanical devices exploiting the SOI technology for a large spectrum of applications, ranging from communications and metrology to sensing and quantum state control.

## References

[1] A. Dorsel, J. D. McCullen, P. Meystre, E. Vignes, and H. Walther, "Optical Bistability and Mirror Confinement Induced by Radiation Pressure," Phys. Rev. Lett. **51**(17), 1550–1553 (1983).
[2] M. Eichenfield, J. Chan, R. M. Camacho, K. J. Vahala, and O. Painter, "Optomechanical crystals," Nature **462**, 78 (2009).
[3] M. Aspelmeyer, T. J. Kippenberg, and F. Marquardt, "Cavity optomechanics," Rev. Mod. Phys. **86**(4), 1391-1452 (2014).
[4] G. P. Agrawal, Nonlinear Fiber Optics, 5th ed., Rochester: New York, 2013, pp. 1–21.
[5] R. V. Laer, R. Baets, and D. Van Thourhout, "Unifying Brillouin scattering and cavity optomechanics," Phys. Rev. A **93**, 053828 (2016).
[6] A. H. Safavi-Naeini, D. Van Thoughout, R. Baets, and R. V. Laer, "Controlling phonons and photons at the wavelength scale integrated photonics meets integrated phononics," Optica **6**, 213 (2019).
[7] S. Gundavarapu, G. M. Brodnik, M. Puckett, T. Huffman, D. Bose, R. Behunin, J. Wu., T. Qiu, C. Pinho, N. Chauhan, J. Nohava, P. T. Rakich, K. D. Nelson, M. Salit, and D. J. Blumenthal, "Sub-hertz fundamental linewidth photonic integrated Brillouin laser," Nat. Photon. **13**(1), 60-67 (2019).
[8] B. J. Eggleton, C. G. Poulton, P. T. Rakich, M. J. Steel, and G. Bahl, "Brillouin integrated photonics," Nat. Photon. **13**, 664 (2019).
[9] G. S. Wiederhecker, P. Dainese, and T. P. M. Alegre, "Brillouin optomechanics in nanophotonic structures," APL Photon. **4**, 071101 (2019).
[10] R. Pant, C. G. Poulton, D. Choi, H. Mcfarlane, S. Hile, E. Li, L. Thevenaz, B. Luther-Davies, S. J. Madden, and B. J. Eggleto, "On-chip stimulated Brillouin scattering," Opt. Exp. **19**, 8285 (2011).
[11] G. Arregui, N. D. Lanzillotti-Kimura, C. M. Sotomayor-Torres, and P. D. García, "Anderson Photon-Phonon Colocalization in Certain Random Superlattices," Phys. Rev. Lett. **122**, 043903 (2019).
[12] F. Kargar, B. Debnath, J.-P. Kakko, A. Säynätjoki, H. Lipsanen, D. L. Nika, R. K. Lake, and A. A. Balandin, "Direct observation of confined acoustic phonon polarization branches in free-standing semiconductor nanowires," Nat. Comm. **7**, 13400 (2016).
[13] F.R. Lamberti, Q. Yao, L. Lanco, D. T. Nguyen, M. Esmann, A. Fainstein, P. Sesin, S. Anguiano, V. Villafañe, A. Bruchhausen, P. Senellart, I. Favero, and N. D. Lanzillotti-Kimura, "Optomechanical properties of GaAs/AlAs micropillar resonators operating in the 18 GHz range," Opt. Exp. **25**, 24437 (2017).
[14] M. Esmann, F. R. Lamberti, A. Harouri, L. Lanco, I. Sagnes, I. Favero, G. Aubin, C. Gomez-Carbonell, A. Lemaître, O. Krebs, P. Senellart, and N. D. Lanzillotti-Kimura, "Brillouin scattering in hybrid optophononic Bragg micropillar resonators at 300 GHz," Optica **6**, 854 (2019).
[15] W. Qiu, P. T. Rakich, H. Shin, H. Dong, M. Soljačić, and Z. Wang, "Stimulated Brillouin scattering in nanoscale silicon step-index waveguides," Opt. Exp. **21**, 31402 (2013).
[16] R. Van Laer, B. Kuyken, D. V. Thourhout, and R. Baets, "Interaction between light and highly confined hypersound in a silicon photonic nanowire," Nat. Photon. **9**, 199 (2015).
[17] E. A. Kittlaus, H. Shin and P. T. Rakich, "Large Brillouin amplification in silicon," Nat. Photon. **10**, 463 (2015).
[18] N. T. Otterstrom, R. O. Behunin, E. A. Kittlaus, Z. Wang, and P. T. Rakich, "A Silicon Brillouin laser," Science **360**, 1113 (2016).
[19] J. Chan, A. H. Safavi-Naeini, J. T. Hill, S. Meenehan, and O. Painter, "Optimized optomechanical crystal cavity with acoustic radiation shield," Appl. Phys. Lett. **101**, 081115 (2012).
[20] R. Zhang, J. Sun, "Design of Silicon Phoxonic Crystal Waveguides for Slow Light Enhanced Forward Stimulated Brillouin Scattering," J. of Light. Tech. **35**, 14 (2017).

[21] M. K. Schmid, C. G. Poulton, G. Z. Mashanovich, G. T. Reed, B. J. Eggleton, and M. J. Steel, "Suspended mid-infrared waveguides for Stimulated Brillouin Scattering," Opt. Exp. **27**, 4976 (2019).

[22] E. A. Kittlaus, N. T. Otterstrom, P. Kharel, S. Gertler and P. T. Rakich, "Non-reciprocal interband Brillouin modulation," Nat. Photon. **12**, 613 (2018).

[23] C. J. Sarabalis, Yanni D. Dahmani, R. N. Patel, J. T. Hill, and A. H. Safavi-Naeini, "Release-free silicon-on-insulator cavity optomechanics," Optica **1**(9), 1147-1149 (2017).

[24] C. J. Sarabalis, J. T. Hill, and A. H. Safavi-Naeini, "Guided acoustic and optical waves in silicon-on-insulator for Brillouin scattering and optomechanics," APL Photon. **1**, 071301 (2016).s

[25] J. Doster, S. Hoenl1, H. Lorenz, P. Paulitschke, and E.M. Weig, "Collective dynamics of strain-coupled nanomechanical pillar resonators," Nature communications **10** (1), 1 (2019).

[26] Mohammed Al Lethawe, Mahmoud Addouche, Sarah Benchabane, Vincent Laude,a and Abdelkrim Khelif, "Guidance of surface elastic waves along a linear chain of pillars" AIP ADVANCES **6**, 121708 (2016).

[27] C. A. Balanis, Antenna Theory: Analysis and Design, Wiley–Blackwell (2005).

[28] R. Halir, P. J. Bock, P. Cheben, A. Ortega-Moñux, C. Alonso-Ramos, J. H. Schmid, J. Lapointe, D.-X. Xu, J. G. Wangüemert-Pérez, I. Molina-Fernández, S. Janz, "Waveguide sub-wavelength structures a review of principles and applications," Las. Photon. Rev. **9**, 25 (2015).

[29] P. Cheben, R. Halir, J. H. Schmid, H. A. Atwater, D. R. Smith, "Subwavelength integrated photonics," Nature 560, 565 (2018).

[30] Q. Quan, and M. Loncar, "Deterministic design of wavelength scale,ultra-high Q photonic crystal nanobeam cavities," Opt. Express **19**, 18529 (2011).

[31] P. T. Rakich, C. Reinke, R. Camacho, P. Davids, and Z. Wang, "Giant Enhancement of Stimulated Brillouin Scattering in the Subwavelength Limit," Phys. Rev. X **2**, 011008 (2012).

[32] P. T. Rakich, P. Davids, and Z. Wang, "Tailoring optical forces in waveguides through radiation pressure and electrostrictive forces," Opt. Exp. **18**, 14439 (2010).

[33] C. Wolff, M. J. Steel, B. J. Eggleton and C. G. Poulton, "Stimulated Brillouin scattering in integrated photonic waveguides Forces, scattering mechanisms and coupled-mode analysis," Phys. Rev. A **92**, 013836 (2015).

[34] Steven G. Johnson, M. Ibanescu, M. A. Skorobogatiy, O. Weisberg, J. D. Joannopoulos, and Y. Fink, "Perturbation theory for Maxwell's equations with shifting material boundaries," Phys. Rev. A **65**, 066611 (2002).